\documentclass[a4paper,12pt]{article}
\pdfoutput=1

\makeatletter

\@addtoreset{equation}{section}
\makeatother
\usepackage{amsfonts}
\usepackage{amsmath}
\usepackage{latexsym}

\usepackage{mathrsfs}

\def\be{\begin{equation}}
\def\ee{\end{equation}}
\def\bea{\begin{eqnarray}}
\def\eea{\end{eqnarray}}
\def\({\left(}
\def\){\right)}
\def\<{\left<}
\def\>{\right>}

\def\[{\left[}
\def\]{\right]}

\def\be{\begin{equation}}
\def\ee{\end{equation}}
\def\bea{\begin{eqnarray}}
\def\eea{\end{eqnarray}}

\def\({\left(}
\def\){\right)}
\def\<{\left<}
\def\>{\right>}

\def\be{\begin{equation}}
\def\ee{\end{equation}}
\def\bea{\begin{eqnarray*}}
\def\eea{\end{eqnarray*}}
\def\ben{\begin{eqnarray}}
\def\een{\end{eqnarray}}
\def\({\left(}
\def\){\right)}
\def\<{\left<}
\def\>{\right>}
\def\!{\right|}
\def\|{\left|}

\def\[{\left[}
\def\]{\right]}

\def\+{\bar}
\def\mb{\mathbb}

\def\Vol{{\mbox{Vol}}}

\def\M{{\cal{M}}}

\def\O{{\cal{O}}}

\def\eps{{\cal{\varepsilon}}}

\def\l{{{\ell}}}

\def\h{\widehat}

\begin{document}

\pagestyle{empty}
\vskip-10pt
\vskip-10pt
\begin{center}
\vskip 3truecm
{\Large\bf
Nonabelian M5-brane on $S^6_q$}
\vskip1cm \centerline{ \textsc{
 Dongsu Bak$^{\, \tt a,b}$ and Andreas Gustavsson$^{\, \tt a}$ } }
\vspace{1cm} 
\centerline{\sl  a) Physics Department,
University of Seoul,  
Seoul 02504 \rm KOREA}
   \vskip0.3cm
 \centerline{\sl b) Natural Science Research Institute,
University of Seoul, 
Seoul 02504 \rm KOREA}
 \vskip0.4cm
\begin{center}
(\tt dsbak@uos.ac.kr,\,agbrev@gmail.com)
\end{center}
\end{center}
\vskip 2truecm
{\abstract We compute the conformal anomaly of a nonabelian M5 brane on $S^1_q\times H^5$ in the large $N$ limit by using the gravity dual of a black hole. We also obtain a general formula for this conformal anomaly for any gauge group by combining various results already present in the literature. From the conformal anomaly we extract the Casimir energy on $\mb{R} \times S^5$. We find agreement with the proposal in arXiv:1507.08553.}

\vfill
\vskip4pt
\eject
\pagestyle{plain}

\section{Introduction}
The six-dimensional nonabelian tensor multiplet theory has no short name. In this paper we will therefore use the name nonabelian M5 brane. To count the number of supersymmetries the gauge group does not play any role and we can take it to be abelian. Then we will refer to a theory with the field content of the abelian 6d (2,0) tensor multiplet in flat space, as the M5 brane. If we put this M5 brane on a curved manifold we may reduce the amount of supersymmetry. But as long as the field content does not change, we will still call this an M5 brane. When we take the gauge group to be nonabelian, for the same geometry and with the same amount of supersymmetry, we will simply refer to that theory as the nonabelian M5 brane, even if a brane realization in M-theory is lacking for some particular gauge group. 

The nonabelian gauge group has three important numbers associated with it, the rank $r$, the dimension $d$ and the dual Coxeter number $h^{\vee}$. For $SU(N)$ gauge group these numbers are
\ben
r &=& N-1\cr
d &=& N^2 - 1\cr
h^{\vee} &=& N\label{sun}
\een
Despite there are some mysteries regarding how to define the nonabelian M5 brane, there are by now a few explicit results available for the nonabelian M5 brane in the literature. For all these quantities, the dependence on the gauge group is of the form 
\bea
{\cal O} &=& r {\cal O}_{U(1)} + d h^{\vee} {\cal O}'  
\eea
where ${\cal O}_{U(1)}$ is the corresponding quantity for abelian gauge group. Intuitively, this term can be understood as arising from breaking the gauge group down to its maximal torus $U(1)^r$ by giving a generic vacuum expectation value to one of the scalar fields leading to a theory of $r$ free tensor multiplets. The second term should then arise from some sort of interactions among these tensor multiplets \cite{Intriligator:2000eq, Bolognesi:2011rq}.

The first nonabelian quantity to be computed was the anomaly 8-form polynomial
\ben
I_8 &=& r I_{8,U(1)} + d h^{\vee} \frac{p_2(N)}{24}\label{eight}
\een
for the nonabelian M5 brane, where $I_{8,U(1)}$ is the anomaly polynomial for the abelian M5 brane \cite{Witten:1996hc} and $p_2(N)$ is a certain invariant 8-form that is computed from the curvature associated with the normal bundle of the M5 brane, which is the same 8-form regardless the choice gauge group. All the nonabelian structure sits in the coefficients $r$ and $d h^{\vee}$. This result was derived for $SU(N)$ and $SO(N)$ gauge groups in \cite{Harvey:1998bx, Yi:2001bz} by using M5 brane embedding into M theory, and its form was conjectured for the $E_r$ gauge algebras in \cite{Intriligator:2000eq}.

A proposal for the Casimir energy for the nonabelian M5 brane on $\mb{R}\times S^5$ was presented in \cite{Bobev:2015kza}. The formula that they presented for the Casimir energy reads
\ben
E_C &=& r E_{U(1)} - d h^{\vee} \frac{\sigma^2_1 \sigma_2^2}{24}\label{Bull}
\een
Here $E_{U(1)}$ denotes the Casimir energy for the abelian M5 brane on $\mb{R}\times S^5$, and 
\bea
\sigma_1 &=& \frac{3}{2} - m\cr
\sigma_2 &=& \frac{3}{2} + m
\eea
where $m$ is a parameter that parametrizes the hypermultiplet mass, such that $m=\pm 1/2$ gives enhanced supersymmetry with $16$ supercharges. The formula (\ref{Bull}) was obtained in \cite{Bobev:2015kza} by using a certain recipe. This recipe says that one shall replace the 8-form anomaly polynomial (\ref{eight}) with its equivariant counterpart that contains forms of all degrees. Then one shall integrate that equivariant form over the M5 brane worldvolume to get a number, and it is this number that is claimed to be the Casimir energy. In practice the integration over the M5 brane worldvolume is done using the index theorem by Berline and Vergne that reduces the integral to a simple evaluation at fixed points. In this application there will be just one fixed point located at the origin. A motivation for this recipe seems to be lacking so far. Perhaps such a motivation can be found by using the relation between the conformal anomaly and the Casimir energy in \cite{Herzog:2013ed}.

The nonabelian conformal anomaly $a$-coefficient were computed in \cite{Cordova:2015vwa, Maxfield:2012aw}. The conformal anomaly $c$-coefficient was conjectured in \cite{Beem:2014kka}. These results are
\ben
a &=& r a_{U(1)} + \frac{16}{7} d h^{\vee} a_{U(1)}\label{aanom}\\
c &=& r c_{U(1)} + 4 d h^{\vee} c_{U(1)}\label{canom}
\een
respectively. 

Using conjectured and proven results above, the nonabelian Renyi entropy on $S^6_{q}$ or any of its conformally equivalent spaces, was subsequently found in \cite{Zhou:2015kaj}
\bea
S(q) &=& r S(q)_{U(1)} + d h^{\vee} H(q)
\eea
where $S(q)_{U(1)}$ is the abelian Renyi entropy that was computed in \cite{Nian:2015xky, Gustavsson:2019ick} and $H(q)$ is a cubic polynomial in $\gamma=1/q$ whose explicit form was found in \cite{Zhou:2015kaj}. 

To this list of exact results, we will here add the conformal anomaly on $S^6_q$. This is a deformation of $S^6$ that introduces a conical singularity. No general formula is presently known for the conformal anomaly on singular spaces, and our result may be seen as a first step towards finding such a general formula. Our result is that the conformal anomaly on $S^6_q$ is given by 
\bea
a(q) &=& r a_{U(1)}(q) + d h^{\vee} b(q)
\eea
where 
\bea
a_{U(1)}(q,x) &=& \frac{x^2}{12 q^3} + \frac{1}{q^2} \(\frac{x}{12} - \frac{x^3}{3}\) + \frac{1}{q} \(\frac{1}{12} - \frac{x}{12} + \frac{x^2}{2}\) \cr
&& + \frac{5}{12} - \frac{x}{12} - \frac{x^2}{3} + \(\frac{1}{12} + \frac{x}{12} + \frac{x^2}{12}\) q
\eea
is the abelian conformal anomaly, and 
\bea
b(q,x) &=& \frac{x^2}{12 q^3} + \frac{1}{q^2} \(\frac{x}{3}-\frac{x^2}{3}\) + \frac{1}{q} \(\frac{1}{3}-\frac{x}{3}+\frac{x^2}{2}\) \cr
&& + \frac{2}{3} - \frac{x}{3} - \frac{x^2}{3} + \(\frac{1}{3} + \frac{x}{3} + \frac{x^2}{12}\) q
\eea
Here 
\bea
x &=& \frac{1}{4} - m^2
\eea
We have $x=0$ when $m=\pm 1/2$. These are the points where we have enhanced supersymmetry with 16 supercharges. For $x=0$ we get 
\bea
a_{U(1)}(q,0) &=& \frac{1}{12 q} + \frac{5}{12} + \frac{q}{12}\cr
b(q,0) &=& \frac{1}{3q} + \frac{2}{3} + \frac{q}{3} 
\eea
When $q=1$ we have the undeformed $S^6$ and we reproduce the known conformal anomaly \cite{Cordova:2015vwa, Maxfield:2012aw, Gustavsson:2019efu,Henningson:1998gx}
\bea
a(1) &=& r\frac{7}{12} + d h^{\vee} \frac{4}{3}
\eea 
This result is consistent with (\ref{aanom}) if one takes $a_{U(1)} = 7/12$. The conformal anomaly is $a(1) = a_{U(1)} = 7/12$ on $S^6$ when we have a single abelian M5 brane on $S^6$.

We also extract the Casimir energy on $S^5$ from the conformal anomaly on $S^6_q$ and we find agreement with (\ref{Bull}). We also obtain the Casimir energy on $S^5$ from supergravity and again find agreement with (\ref{Bull}) for $SU(N)$ gauge group in the large $N$ limit. 

The partition function on $S^5$ with radius $r$ (not to be confused with the rank $r$ of the gauge group) has been computed in \cite{Minahan:2013jwa} (see also \cite{Kim:2012ava}) for $SU(N)$ gauge group by using localization of 5d SYM with Yang-Mills coupling constant $g_{YM}$. The result that was obtained by taking the large $N$ limit of the resulting matrix model, was the following logarithm of the partition function,
\ben
- \ln Z &=& - \frac{g_{YM}^2 N^3}{96\pi r} \(\frac{9}{4} - m^2\)^2 + {\cal O}(N^2)\label{Mina1}
\een
The partition function is determined by the Casimir energy $E_C(N)$ in the limit $\beta \rightarrow \infty$ as
\bea
Z &=& e^{-\beta E_C(N)}
\eea
Let us now apply (\ref{Bull}) with the gauge group $SU(N)$ and take the large $N$ limit. We then get
\ben
E_C(N) &=& - \frac{N^3}{24 r}  \(\frac{9}{4} - m^2\)^2 + {\cal O}(N^2)\label{BullN}
\een
If we assume that the relation between the 5d SYM coupling constant and the radius of the circle on which the M5 brane is compactified is given by 
\ben
g_{YM}^2 &=& 4 \pi^2 \beta\label{Lam}
\een
then we get 
\ben
\beta E_C(N) &=& - \frac{g_{YM}^2 N^3}{96\pi r} \(\frac{9}{4} - m^2\)^2 + {\cal O}(N^2)\label{Bull1}
\een
We now see that (\ref{Bull1}) agrees with (\ref{Mina1}), which shows that the relation (\ref{Lam}) is valid. This, however, is not apriori obvious. Once we compactify the M5 brane on $S^1 \times S^5$, we have a notion of strong and weak coupling regimes because we have a dimensionless `t Hooft parameter $\lambda = g_{YM}^2 N/r$ where $r$ is the radius of $S^5$, and $\beta$ is the circumference of $S^1$. It is not obvious that the same relation (\ref{Lam}) holds in both the strongly coupled regime where $\lambda \gg 1$ as in the weakly coupled regime where $1 \gg \lambda $. In the decompactification limit $r\rightarrow \infty$ there is only the weakly coupled regime and in that case we may expect (\ref{Lam}) to be true for all values of $\beta$. When $r$ is finite, this is no longer obvious. But the agreement between (\ref{Mina1}) and (\ref{Bull1}) suggests that (\ref{Lam}) is valid also in the strongly coupled regime $\lambda \gg 1$. The authors in \cite{Minahan:2013jwa} did not know about the result (\ref{Bull1}) which came later, and they came to a different conclusion by studying the gravity dual side. They found that (\ref{Lam}) should be replaced by a different relation for $\lambda \gg 1$. However, the gravity dual of the geometry $S^1 \times S^5$ is not known when the R gauge fields are turned on as is necessary to preserve some amount of supersymmetry. 

We will approach the problem in an indirect way, by following the approach in \cite{Zhou:2015kaj}. We begin with $S^6_q$, which is conformally equivalent with $S^1_q \times H^5$. For M5 brane on this space we have a smooth gravity dual, which is a two-charged black hole solution \cite{Cvetic:1999xp} where we can turn on R gauge fields and preserve supersymmetry. By using this gravity dual, we get the conformal anomaly on $S^6_q$ and the Casimir energy on $H^5$. Then we will use a correspondence that enables us to extract the Casimir energy on $S^5$ from the conformal anomaly on $S^6_q$. We find agreement with (\ref{BullN}) in the large $N$ limit.

\section{A correspondence between $S^d_q$ and $\mb{R}\times S^{d-1}$}\label{relation}
It has been noticed \cite{Beccaria:2017lcz,Gustavsson:2019ick} that the general structure of the conformal anomaly always appears to be on the form 
\bea
a_d(q) &=& \frac{\nu}{f(d) q^{d-1}} + ... + (-2 r E_C) q
\eea
for any conformal field that one puts on $S^d_q$ in any dimension $d$. Here $\nu$ is the number of degrees of freedom, and $E_C$ is the corresponding Casimir energy that one would obtain by putting the theory instead on $\mb{R}\times S^{d-1}$ where $r$ is the radius of $S^{d-1}$. The function $f(d)$ is some function of the dimension, but its precise form is unknown.

Let us first show the validity of this correspondence for a single conformal scalar in $d$ dimensions. It has the euclidean action
\bea
S &=& \frac{1}{2} \int d^d x \sqrt{g} \(g^{\mu\nu} \partial_{\mu}\phi \partial_{\nu} \phi + \frac{d-2}{4(d-1)} R \phi\)
\eea
Let us now put this scalar on $S^d_q$ where we have $R = d(d-1)$ and we put $r=1$ for simplicity. Then the action becomes
\bea
S &=& \frac{1}{2} \int d^d x \sqrt{g} \phi \(\triangle + \frac{d(d-4)}{4}\) \phi
\eea
We have the eigenvalues 
\bea
\triangle &=& (n+\gamma m)(n+\gamma m + d-1)
\eea
and so for the conformal scalar we get
\bea
\triangle + \frac{d(d-4)}{4r^2} &=& \(n+\gamma m + \frac{d-2}{2}\) \(n+\gamma m + \frac{d}{2}\)
\eea
The corresponding half heat kernels are 
\bea
K_{d,1}(t) &=& \sinh \frac{\gamma t}{2} \sum_{n=0}^{\infty} d_{n,0}  e^{-t\(n+\frac{d-2}{2}\)}\cr
K_{d,2}(t) &=& \sinh \frac{\gamma t}{2} \sum_{n=0}^{\infty} d_{n,0}  e^{-t\(n+\frac{d}{2}\)}
\eea
where 
\bea
d_{n,0} &=& \(\begin{array}{c}
n+d-2\\
n
\end{array}\)
\eea
For an explanation of these half heat kernels, we refer to \cite{Beccaria:2017lcz, Gustavsson:2019ick}. Here it will suffice to know that they serve as a tool to compute the conformal anomaly. We expand the average sum of these half heat kernels for small $t$ and find a series expansion of the form 
\bea
K_d(t) = \frac{1}{2} \(K_{d,1}(t)+K_{d,2}(t)\) = \frac{2}{t^6} + ... + a_d + {\cal O}(t)
\eea
From this expansion, we can extract the conformal anomaly $a_d$. We find the following results
\bea
a_6 &=& \frac{1}{15120 q^5} + \frac{1}{4320 q^3} + \frac{31 q}{30240}\cr
a_5 &=& 0\cr
a_4 &=& - \frac{1}{360 q^3} - \frac{q}{120}\cr
a_3 &=& 0\cr
a_2 &=& \frac{1}{6q} + \frac{q}{6}\cr
a_1 &=& 0
\eea

On the other side of the correspondence, there are the Casimir energies on $S^D = S^{d-1}$ of a conformal scalar, which we may extract from the single particle index. Here we of course have no parameter $q$, and the correspondence says that it is the linear term in $q$ that corresponds to the Casimir energy on $S^D$. We will not confirm this by explicitly computing the Casimir energies for the first few values of $d$. The single particle index is given by
\bea
f_d(\eps) &=& \sum_{n=0}^{\infty} d_{D,n} e^{-\eps \(n+\frac{D-2}{2}\)}
\eea
where the degeneracy of spherical harmonics on $S^D$ is given by
\bea
d_{D,n} &=& \(\begin{array}{c}
D+n\\
n
\end{array}\) - \(\begin{array}{c}
D+n-2\\
n-2
\end{array}\)
\eea
The Casimir energy is given by 
\bea
-2E_d &=& f'_{d,ren}(0)
\eea
where $ren$ means that we subtract the divergences before we take $\eps = 0$. The results we get are, if we also restore the dependence on $r$,
\bea
-2E_6 &=& \frac{31}{30240 r}\cr
-2E_5 &=& 0\cr
-2E_4 &=& -\frac{1}{120 r}\cr
-2E_3 &=& 0\cr
-2E_2 &=& \frac{1}{6 r}\cr
-2E_1 &=& 0
\eea
We checked the correspondence also for $S^{14}$ and $\mb{R}\times S^{13}$ and got agreement,
\bea
a_{14} &=& \frac{1}{37362124800q^{13}} + ... + \frac{3203699q}{3138418483200}\cr
-2E_C &=& \frac{3203699}{3138418483200 r}
\eea
so we may expect the correspondence holds in any dimension. 

Supersymmetry does not play any part in this correspondence. It apparently holds for a single conformal scalar field. Instead it is conformal symmetry that plays the key role in this correspondence. This can be easily seen. If we drop the conformal mass term from the conformal scalar field theory, this correspondence no longer holds. While it is easy to see that $S^d_q$ is conformally equivalent to $\mb{R} \times S^{d-1}_q$, we have not been able to find any conformal map that would relate this to $S^1_q \times S^{d-1}$. Instead it is conformally equivalent to $S^1_q \times H^{d-1}$. This fact will provide us with a useful relation between the Casimir energies on $H^{d-1}$ and $S^{d-1}$ that we will use later. The metric on $S^d_q$ can be written as
\ben
ds^2 &=& r^2 \(d\theta^2 + \cos^2 \theta d\Omega_{d-2}^2 + q^2 \sin^2 \theta d\tau^2\)\label{metr}
\een
where $\theta \in [0, \frac{\pi}{2}]$ and $\tau \sim \tau + 2\pi$. 
This metric is conformally equivalent with the metric on $S^1_q \times H^{d-1}$,
\bea
ds^2 &=& r^2 \(q^2 d\tau^2 + d\Xi_{d-1}^2\)\cr
d\Xi_{d-1}^2 &=& d\eta^2 + \sinh^2 \eta d\Omega_{d-2}^2
\eea
The relation between these coordinates is $\cot \theta = \sinh \eta$. Then $\eta \in [0,\infty)$ and the conical singularity at $\theta = 0$ corresponds to $\eta = \infty$. We shall regularize the volume of $H^{d-1}$ by introducing a cutoff at a finite $\eta = \eta_0$ that we shall define such that
\bea
e^{-\eta_0} &=& \frac{\eps}{r}
\eea
for a small cutoff length $\eps$. Here we divide this by the only other length scale in the problem, which is $r$, to get a dimensionless ratio. This cutoff amounts to cutting off the conical singularity. The partition function on $S^d_q$ can be written as
\bea
Z(S^d_q) &=& e^{a_d(q) \ln \(\mu r\)}
\eea
where $\mu$ is some mass scale that we introduce to get a dimensionless combination $\mu r$, and $a_d$ denotes the conformal anomaly as a function of $q$. For a very large $q \gg 1$, the partition function is dominated by the linear term $a_d(q) \sim q (-2 E_C)$ that multiplies the Casimir energy,
\bea
Z(S^d_q) &\sim & e^{-2 q E_C(S^5) \ln \(\mu r\)}
\eea
Let us next turn to the space $S^1_q \times H^5$ and let time run along the $S^1_q$. In that case, when we take $q \gg 1$, the leading behavior is governed by the Casimir energy on $H^{d-1}$,
\bea
Z(S^1_q \times H^{d-1}) &\sim & e^{-2\pi \l q E_C(H^{d-1})}
\eea
Generically the partition function is not conformally invariant because there may be a conformal anomaly. But we may expect that terms that are proportional to a logarithm are conformally invariant as they are part of a conformal anomaly. This leads us to conjecture that the exponents in the above partition functions are equal, which in turn relates the Casimir energies on these spaces as
\bea
E_C(S^{d-1})  &=& \frac{\pi \l}{\ln \(\mu r\)} E_C(H^{d-1})
\eea
From the gravity computation corresponding to $S^1_q \times H^5$ we obtain the result
\bea
E_C(H^5) &=& - \frac{N^3}{24 \pi r} (2+x)^2 \ln \(\frac{r}{\eps}\)
\eea
Using the above relation we then conclude that 
\bea
E_C(S^5) &=& - \frac{N^3}{24 r}(2+x)^2
\eea
if we decide to choose the scales $\mu$ and $\eps$ such that $\mu \eps = 1$ so that the log terms cancel out, which is something we shall expect if the Casimir energy is related to the anomaly, since the anomaly has no log dependent term. This result is in then agreement with (\ref{Bull1}). 

\subsection{A motivation of the correspondence}
We notice that the partition function on $S^d_q$ is given by
\ben
Z(S^d_q) &=& e^{a_d(q) \ln (\mu r)} \label{ge1}
\een
We are now interested in determining the coefficient $a_{-1}$. We will do this by using the partition function on $H^1 \times S^{d-1}_q$ that we write on the form 
\bea
Z(H^1 \times S^{d-1}_q) &=& \exp\(-\Vol(H^1) F(S^{d-1}_q)\)
\eea
where $F(S^{d-1}_q)$ is the free energy on $S^{d-1}_q$. By conformal invariance of the universal part of this partition function, we have 
\ben
\Vol(H^1) F(S^{d-1}_q) &=& - a_d(q) \ln (\mu r)\label{free}
\een
We know that $a_d(q)$ has the general form 
\ben
a_d(q) &=& a_{d-1} q^{-(d-1)} + ... + a_{-1} q\label{ge2}
\een
By equating the left and right hand sides in (\ref{free}) we conclude that the free energy must have the same general form
\bea
F(S^{d-1}_q) &=& F_{d-1} q^{-(d-1)} + ... + F_{-1} q
\eea
By equating the coefficients of the linear terms in $q$, we get the relation
\bea
F_{-1} &=& - \frac{\ln(\mu r)}{\Vol(H^1)} a_{-1}
\eea
By inserting the volume
\bea
\Vol(H^1) &=& 2 r \ln \(\frac{r}{\eps}\)
\eea
we get
\bea
F_{-1} &=& - \frac{a_{-1}}{2}
\eea

To complete the argument, we would like to show that $F_{-1} = E_C$ is the Casimir energy $E_C$ on $S^{d-1}$. If we take the infinite $q$ limit of $S^d_q$, we find the result \cite{Beccaria:2017lcz}\footnote{\label{D}This is not correct near the conical singularity where the local geometry is a conical disk times a sphere $D^2_q \times S^{d-2}$. If we  cut out the tip of the cone from $S^6_q$ then this will correspond to replacing $\mb{R} \times S^{d-1}$ with an open manifold $\mb{R} \times D^{d-1}$ whose boundary is $\mb{R} \times S^{d-2}$ near the tip of the cone. This boundary corresponds to the boundary of $D^2_q \times S^{d-2}$ that we cut out, in the limit $q\rightarrow \infty$. In footnote \ref{D1} we will argue that physics on $\mb{R} \times D^{d-1}$ with Dirichlet boundary condition approaches physics on $\mb{R} \times S^{d-1}$ in a smooth manner, as we let the boundary approach the tip of the cone. Therefore we will be sloppy about distinguishing between $D^{d-1}$ and $S^{d-1}$ from now on.}
\bea
\lim_{q\rightarrow \infty} S^d_q &=& \mb{R} \times S^{d-1}
\eea
To see this, we define a new time variable as 
\bea
t &=& r q \tau \sin \theta
\eea
and change the ranges of coordinates so that $\tau \in [0,\pi]$ and $\theta \in [-\pi/2,\pi/2]$. Then we may use the approximation  
\bea
dt \approx r q d \tau \sin \theta
\eea
to get the approximate metric
\bea
ds^2 &=& dt^2 + r^2 d\Omega_{d-1}^2
\eea
on $S^d_q$ as long as 
\ben
\tau \cos \theta << |\sin \theta|\label{cond}
\een
That is, as long as the range of $\tau$ is sufficiently small. To allow for a finite but small range of $\tau$, we introduce a small cutoff $\theta_0>0$ and let $\tau$ range in the interval $\tau \in [0,\tau_0]$ where we take
\bea
\tau_0 &=& \frac{1}{n} \tan \theta_0
\eea
for some large integer $n>>1$. Then the condition (\ref{cond}) will be met for all $\theta \in I_0 = [-\pi/2,-\theta_0]\cup [\theta_0,\pi/2]$. In this interval we may to a good approximation use the time variable $t$ instead of $\tau$. We may evolve $t$ in the interval $t\in [0,\beta_0]$ where 
\bea
\beta_0 = r q \tau_0 \sin \theta_0 = \frac{1}{n} r q \tan \theta_0 \sin \theta_0
\eea
while $\theta \in I_0$ and let $t$ stay fixed at $t=0$ when $\theta$ is outside $I_0$, which means that we restrict the dynamics to happen only inside $I_0$.\footnote{\label{D1}This is a harmless restriction since from the metric (\ref{metr}) we see that at the tip of the cone where $\theta = 0$, there is no time evolution anyway and so we have the Dirichlet boundary condition $\phi(t,$tip$) = \phi($tip$)$ at the tip of the cone. This is because the term $r^2 q^2 \sin^2 \theta d\tau^2$ in the metric is zero at the tip, which corresponds to the infinite mass limit at the tip. What we do here is that we simply move this boundary condition to some infinitesimally small $\theta_0>0$ away from the tip. By taking the limit $\theta_0\rightarrow 0$ we recover the boundary condition and the partition function on $S^d_q$.}  In this time interval the condition (\ref{cond}) is satisfied and $t$ is a good approximate time variable to use. We may now take $q>>n$ such that $\beta_0>>r$. In this case we will have the dominant contribution coming from the Casimir energy $E_C$ on $S^{d-1}$ to the partition function,
\bea
Z(\beta_0) &\approx & e^{-\beta_0 E_C}
\eea
This is not the partition function on the full $S^6_q$ since we have evolved in the time $t$ such that we covered only a very small portion of $S^6_q$. However, if we evolve further in time nothing essential will happen to the partition function. It will remain of the same form as a function of the time interval and the Casimir energy for all future times to a very good approximation. The only thing that will happen is that a multiplicative constant $c' >1$ will enter the partition function on the full $S^6_q$ as 
\ben
Z(S^d_q) &\approx & e^{-c' \beta_0 E_C}\label{end}
\een
If we were to do a detailed computation to reach this same conclusion, we would need to change our definition of time $t$ as we would continue to evolve in time beyond $\beta_0$ in order to maintain a good approximation at all times. But on general grounds as long as we evolve using a time variable that corresponds to a Killing vector field on $S^6_q$ we should end up with (\ref{end}) for some constant $c'$. Since $\beta_0 \sim q$, we can from this analysis infer that the general dependence on $q$ should be on the form 
\ben
Z(S^d_q) &\approx& \exp\(- c q E_C\)\label{largeq}
\een
for some constant $c$ and for a very large $q$. Here, although $\beta_0$ and $c'$ may depend on the cutoff $\theta_0$, we do not expect $c$ to depend on this cutoff. The relation (\ref{largeq}) will be modified when $q$ is close to $q=1$ by other terms of the general form (\ref{ge1}) and (\ref{ge2}) also giving a significant contribution at the same order (when $q\approx 1$), but it will be still true that the term that is linear in $q$ will have a coefficient that is proportional to the Casimir energy because we know the general structure (\ref{ge1}) and (\ref{ge2}) is valid for all $q$ since that is an exact expression. Thus we may take $q$ close to $q=1$ and extract the Casimir energy from the linear term in $q$. On the other hand, we have by conformal invariance the result 
\ben
Z(S^d_q) &=& \exp\(... - q \Vol(H^1) F_{-1}\)\label{F}
\een
and we know that the time interval of $H^1 \times S^{d-1}$ is $\Vol(H^1)$ for $q=1$. This result we can now apply to (\ref{largeq}) by letting $q\approx 1$ there to conclude that $c q$, which we identify with the time interval, must be given by $c = \Vol(H^1)$ when $q=1$. But since $c$ and $\Vol(H^1)$ are both constants independent of $q$ we conclude that $c=\Vol(H^1)$ must hold for any $q$. By then  identifying the linear term in $q$ in (\ref{F}) and (\ref{largeq}) we conclude that $F_{-1} = E_C$. We have now shown the correspondence
\ben
E_C &=& - \frac{1}{2r} a_{-1}\label{corr}
\een

\section{Gravity computation}
When the gauge group is $SU(N)$ we can take the large $N$ limit where we have a gravity dual description of the M5 brane theory on $S^1_q \times H^5$ which is a certain black hole geometry. Before turning to this black hole solution, let us describe the bulk geometry that has the boundary $S^1_{q=1} \times H^{d-1}$ where in our case $d=6$. The metric on $H^{d+1}$ can be chosen as
\bea
ds^2 &=& \l^2 \(d\rho^2 + \sinh^2\rho d\Omega_{d-k}^2 + \cosh^2 \rho d\Xi_k^2\)
\eea
for any of the values $k=0,1,...,d$. The boundary surface at some large cutoff $\rho = \rho_0$ is $S^{d-k} \times H^k$ and these boundary manifolds for all values of $k$ are conformally equivalent. If we take $k=d-1$ we get the boundary $S^1 \times H^{d-1}$. The metric on $H^{d+1}$ can then be written as
\bea
ds^2 &=& \l^2 \(d\rho^2 + \sinh^2 \rho d\tau^2 + \cosh^2 \rho d\Xi_{d-1}^2\)
\eea
where $\tau \sim \tau + 2\pi$. We may define a radius coordinate as
\bea
r &=& \l \cosh \rho
\eea
in terms of which this metric becomes
\bea
ds^2 &=& \frac{dr^2}{f} + f \l^2 d\tau^2 + r^2 d\Xi_{d-1}^2
\eea
where
\bea
f &=& \frac{r^2}{\l^2} - 1
\eea
We may define a time coordinate as
\bea
t &=& i f \tau 
\eea
that brings this metric into the form 
\bea
ds^2 &=& - f dt^2 + \frac{dr^2}{f} + r^2 d\Xi_{d-1}^2
\eea
where $t \sim  t + 2 \pi i \l$. This corresponds to the inverse temperature being 
\bea
\beta &=& 2\pi \l
\eea
We are now want to have a gravity solution for the case where the inverse temperature is $q$-deformed to 
\bea
\beta &=& 2\pi \l q
\eea
This $q$-deformation corresponds on the gravity side to deforming the $AdS_7$ geometry into a black hole geometry \cite{Cvetic:1999xp} whose metric is given by
\bea
ds^2 &=& - (H_1 H_2)^{-4/5} f dt^2 + (H_1 H_2)^{1/5} \(\frac{dr^2}{f} + r^2 d\Xi_5^2\) 
\eea
where
\bea
f &=& \frac{r^2}{\l^2} H_1 H_2 - 1 - \frac{m}{r^4} \cr
H_i &=& 1 + \frac{q_i}{r^4}\cr
\eea
for $i=1,2$. Here 
\bea
d\Xi_5^2 &=& G_{ij} dx^i dx^j
\eea
is used to denote the metric on $H^5$ with unit radius, whose Riemann curvatures are 
\bea
R_{ij} &=& - (d-2) G_{ij}\cr
R &=& - (d-2)(d-1)
\eea
with $d=6$. The black hole has two electric charges $Q_i$ for $i=1,2$. The corresponding gauge potentials are given by
\bea
A_i &=& \(i \(\frac{1}{H_i}-1\) + \mu_i\) dt
\eea
where $i = 1,2$ labels two Cartan generators of $SO(5)$. This appears as the gauge group of 7d supergravity, it the isometry group of $S^4$, and it appears as the global R-symmetry group of the M5 brane theory. 

The gravity solution that gives $\beta = 2\pi \l q$ has $m = 0$ and $q_i$ are related to $r_i$ as below that has to satisfy the supersymmetric constraint $r_1+r_2=0$, but for the time being we will keep $m$ as well as $r_1$ and $r_2$ arbitrary since the black hole solution exists anyway.

We begin by computing the mass of this black hole by following the procedure of \cite{Balasubramanian:1999re} who compute a quasilocal stress tensor that lives at the boundary. We begin by introducing a notation where the black hole metric takes the form
\bea
ds^2 &=& - A^2 dt^2 + B^2 d\Xi_5^2 + \frac{dr^2}{C^2}
\eea
Here we define
\bea
A &=& (H_1 H_2)^{-2/5} \sqrt{f}\cr
B &=& (H_1 H_2)^{1/10} r\cr
C &=& (H_1 H_2)^{-1/10} \sqrt{f}
\eea
We choose a cutoff boundary surface at $r=r_0$ where we have the boundary metric of $S^1_{q} \times H^5$,
\bea
ds^2_{bndry} &=& \gamma_{\mu\nu} dx^\mu dx^\nu
\eea
whose components are 
\bea
\gamma_{tt} &=& - A^2\cr
\gamma_{ij} &=& B G_{ij}
\eea
The unit normalized normal vector to this surface has the only nonvanishing component
\bea
n_r &=& \frac{1}{C}
\eea
We may introduce a boundary time $T = A t$ in which the boundary metric takes the form
\bea
ds^2_{bndry} &=& - dT^2 + \gamma_{ij} dx^i dx^j
\eea 
We have the extrinsic curvature 
\bea
K_{\mu\nu} &=& \frac{1}{2} \(D_\mu n_\nu + D_\nu n_\mu\)\cr
K &=& \gamma^{\mu\nu} K_{\mu\nu} 
\eea
We get
\bea
K_{tt} &=& - C A A'\cr
K_{ij} &=& C B B' G_{ij}
\eea
and
\bea
K &=& C \(\frac{A'}{A} + \frac{5 B'}{B}\)
\eea
The quasilocal stress tensor is given by
\bea
T_{\mu\nu} &=& T^K_{\mu\nu} + T^{ct}_{\mu\nu}
\eea
where 
\bea
T^K_{\mu\nu} &=& - \frac{1}{8\pi G} \(K_{\mu\nu} - \gamma_{\mu\nu} K\)\cr
T^{ct}_{\mu\nu} &=& \frac{2}{\sqrt{-\gamma}} \frac{\delta I_{ct}}{\delta \gamma_{\mu\nu}}
\eea
Here $I_{ct} = \int dT L_{ct}$ where \cite{Emparan:1999pm}
\bea
L_{ct} &=& \frac{r_0^{d-1}}{8\pi G} \int d^{d-1} x \sqrt{G} \cr
&& \(\frac{d-1}{\l} + \frac{\l}{2(d-2)} R + \frac{\l^3}{2(d-4)(d-2)^2} \(R_{\mu\nu}^2 - \frac{d}{4(d-1)} R^2\)\)
\eea
and since this does not involve a kinetic term, the Hamiltonian is simply $M_{ct} = - L_{ct}$. The mass of the black hole as measured by the boundary proper time $T$ is
\bea
M = \int d^5 x \sqrt{G} B^5 T_{TT}
\eea
The integral is over the boundary surface at $r=r_0$. We get
\bea
M &=& \frac{\Vol(H^5)}{8\pi G} \(\M^K + \M^{ct}\)
\eea
where 
\bea
\M^K &=& - 5 C B' B^4\cr
\M^{ct} &=& \frac{5B^5}{\l} \(1 - \frac{\l^2}{2 B^2} - \frac{\l^4}{8 B^4}\) 
\eea
If we expand $\M^K$ and $\M^{ct}$ in powers of the expansion parameter $\delta = \l/r_0$ up to linear order in $\delta$ we get
\bea
\M^{K} &=& -\frac{5 \l^4}{\delta^5} + \frac{5 \l^4}{2 \delta^3} + \frac{5 (\l^4 - 4 q_1 - 4 q_2)}{8 \delta} + 
 \frac{5}{16} (\l^4 + 8 m - 4 q_1 - 4 q_2) \delta + {\cal O}(\delta^2)\cr
\M^{ct} &=& \frac{5 \l^4}{\delta^5} - \frac{5 \l^4}{2 \delta^3} - \frac{5 (\l^4 - 4 q_1 - 4 q_2)}{8 \delta} - 
 \frac{3}{4} (q_1 + q_2) \delta + {\cal O}(\delta^2)
\eea
We now see that the counterterms cancel all the divergent terms, and we are left with a finite mass
\bea
M &=& \frac{\Vol(H^5) \l^5}{8\pi G} \(\frac{5}{16} + \frac{5m}{2\l^4} - \frac{2}{\l^4} \(q_1+q_2\)\) \frac{1}{r_0} 
\eea
If $G$ denotes the 11d gravitational constant, then we shall multiply this result by $\Vol(S^4) = \pi^4 \l^4/6$. We have the following 11d relation \cite{Maldacena:1997re}\footnote{The 11d Newton constant is $G = 16 \pi^7 \l_P^9$ and the radius of $AdS_7$ is related to the rank $N$ of the gauge group of the dual CFT as $\l=\l_P(\pi N)^{1/3}$. We assume that this latter AdS-CFT relation remains intact by the $q$-deformation that deforms $AdS_7$ into a black hole.}
\bea
\frac{\pi^4 \l^9}{G} &=& 32 N^3
\eea
and we have the regularized value for the volume of $H^5$ of unit radius, 
\bea
\Vol(H^5) &=& \pi^2 \ln\frac{{r_0}}{\eps}
\eea
where $\eps$ is a cutoff scale. For a computation of this volume, see Appendix B. We then get the mass 
\bea
M &=& \frac{4N^3}{3} \frac{1}{2\pi r_0} \(\frac{5}{16} + \frac{5m}{2\l^4} - \frac{2}{\l^4}\(q_1+q_2\)\) \ln\frac{{r_0}}{\eps}
\eea
However, this mass may have to be shifted by a constant shift, where the constant does not depend on the parameter $q$. To fix that constant, we notice that the bulk metric can be expanded as 
\bea
g_{MN} &=& g_{MN}^0 + b g_{MN}^1 + {\cal O}(b^2)
\eea
where $b := 1/q - 1$. All the supergravity fields has a corresponding expansion in powers of $b$. The action is on-shell when $b = 0$, which means that the action for a nonzero but small $b$, the action will be on the form $I(b) = I(b=0) + \O(b^2)$. The term that is linear in $b$ when we expand the action vanishes since it multiplies the equations of motion that vanish on-shell. The energy is computed from the action as
\bea
E(b) &=& \frac{\partial I}{\partial \beta} - \frac{\mu_i}{\beta} \frac{\partial I}{\partial \mu_i}
\eea
Both $d\beta$ and $d\mu_i$ are proportional to $db$, so when we evaluate $E(b)$ at $b=0$ we get zero,
\bea
E(0) &=& 0
\eea
This fixes the constant shift of the energy such that the shifted mass becomes
\bea
M &=& - \frac{4N^3}{3} \frac{2}{2\pi r_0} \frac{q_1+q_2}{\l^4} \ln\frac{{r_0}}{\eps}
\eea

This is the mass measured by the proper boundary time $T$ that we define from the boundary metric 
\bea
ds_{bndry}^2 &=& - dT^2 + ...
\eea
However, the other black hole state variables, the entropy and the temperature, are computed using the coordinate time $t$. These are related as $T = A t$. To leading order we have $A = \frac{r_0}{\l}$ so we find that the mass measured in coordinate time $t$ is given by 
\bea
M &=& - \frac{4N^3}{3} \frac{2}{2\pi \l} \frac{q_1+q_2}{\l^4} \ln\frac{{r_0}}{\eps}
\eea
For the black hole, the natural parameter to use is $\kappa_i$ in place of $q_i$. These are related as 
\bea
\kappa_i &=& \frac{q_i}{r_H^4}
\eea
where the horizon radius is
\bea
r_H &=& \frac{\l}{\sqrt{\(1+\kappa_1\)\(1+\kappa_2\)}}
\eea
Expressed in terms of $\kappa_i$, the mass becomes
\bea
M &=& - \frac{4N^3}{3} \frac{1}{\pi \l} \frac{\kappa_1+\kappa_2}{\(1+\kappa_1\)^2 \(1+\kappa_2\)^2} \ln\frac{{r_0}}{\eps}
\eea
\bea
E &=& - \frac{S_0}{\pi \l} \frac{\kappa_1+\kappa_2}{(1+\kappa_1)^2(1+\kappa_2)^2}
\eea

The on-shell gravity action can be expanded in terms of state variables of the black hole as follows, 
\bea
I &=& \beta E - S - \beta \(\mu_1 Q_1 + \mu_2 Q_2\)
\eea
Here
\bea
\beta &=& \frac{1}{T}
\eea
is the inverse temperature of the black hole. The temperature, entropy and charges of the black hole were computed in \cite{Zhou:2015kaj} with the results
\ben
T &=& \frac{1-\kappa_1-\kappa_2-3\kappa_1\kappa_2}{2\pi \l (1+\kappa_1)(1+\kappa_2)}\label{T}\\
S &=& S_0 \frac{1}{(1+\kappa_1)^2 (1+\kappa_2)^2}\\
Q_i &=& \frac{i S_0}{\pi \l} \frac{\kappa_i}{(1+\kappa_1)^2(1+\kappa_2)^2}
\een
Also the chemical potentials $\mu_i$ were determined from demanding regularity of the gauge potential at the black hole horizon \cite{Chamblin:1999hg}
\bea
\mu_i &=& \frac{i \kappa_i}{\kappa_i + 1}
\eea
Here 
\bea
S_0 &=& \frac{4N^3}{3} \ln\frac{{r_0}}{\eps}
\eea
is the entropy at $q=1$. Inserting all this into the on-shell action we get
\bea
I &=& -  \frac{S_0}{(1+\kappa_1)(1+\kappa_2)(1-\kappa_1\kappa_2-3\kappa_1\kappa_2)}
\eea
The chemical potentials were found in \cite{Zhou:2015kaj, Huang:2014gca} to be related to the parameters $r_1$ and $r_2$ of the boundary CFT as
\bea
\mu_i &=& -\frac{i r_i}{2} b
\eea
This gives us
\bea
\kappa_i &=& - \frac{b r_i}{2} \frac{1}{1 + \frac{b r_i}{2}}
\eea
Using this, we can compute the temperature using (\ref{T}) with the result
\bea
T &=& \frac{1+b\(r_1+r_2\)}{2\pi \l}
\eea
We thus again see the necessity to demand that $r_1+r_2=1$. Here this is necessary in order to match with $\beta = 2 \pi \l q$. The on-shell action becomes
\bea
I &=& - S_0 \frac{(2+b r_1)^2 (2+b r_2)^2}{16\(1+b(r_1+r_2)\)}
\eea
Imposing the condition $r_1 + r_2 = 1$, we get
\bea
I &=& - N^3 \frac{\(4 + 2 b + b^2 x\)^2}{12 (1+b)} \ln\frac{{r_0}}{\eps}
\eea
By identifying the exponent of the on-shell gravity action $I$ with the partition function $Z$ of the boundary CFT
\bea
e^{-I} = Z
\eea
we may extract the conformal anomaly of the CFT
\bea
a(q) &=& N^3 \frac{\(4 + 2 b + b^2 x\)^2}{12 (1+b)}
\eea
When $q \gg 1$ the CFT partition function behaves like 
\bea
Z \sim e^{-\beta E_C}
\eea
where $E_C(H^5)$ is the Casimir on $H^5$. From this we get
\bea
E_C(H^5) &=& - \frac{N^3}{24 \pi \l} (2+r_1 r_2)^2 \ln\frac{{r_0}}{\eps}
\eea
Although this Casimir energy depends on three lengths $\l$, $r_0$ and $\eps$ and two arbitrary cutoff scales $\l/r_0$ and $\eps$, it contains precise information about the Casimir energy on $S^5$. First, by measuring this Casimir energy by the boundary CFT time variable $T$, we eliminate the length $\l$,
\bea
E_C(H^5) &=& - \frac{N^3}{24 \pi r_0} (2+r_1 r_2)^2 \ln\frac{{r_0}}{\eps}
\eea
and second, this is related to the Casimir energy on $S^5$ of radius $r_0$ by relations explained in section \ref{relation} 
\bea
E_C(S^5) &=& - \frac{N^3}{24 r_0} (2+r_1 r_2)^2
\eea
where the log dependence has disappeared. We have now got a unique expression for the Casimir energy on $S^5$ with radius $r_0$ in the large $N$ limit as measured by the boundary CFT proper time. Alternatively we get $E_C(S^5)$ directly from the conformal anomaly by extracting the coefficient $a_{-1}$ and by using (\ref{corr}).

\section{The abelian Casimir energy}
The abelian Casimir energy of a single M5 brane on $\mb{R} \times S^5$ was computed in \cite{Bak:2016vpi}. The result is 
\bea
E_C &=& - \frac{7}{128} + \frac{m^2}{16} - \frac{m^4}{24}
\eea
The mass parameter $m$ is related to $r_1$ and $r_2$ as
\bea
r_1 &=& \frac{1}{2}+m\cr
r_2 &=& \frac{1}{2}-m
\eea
Using this, we can write the abelian Casimir energy as
\ben
E_C &=& - \frac{1}{24} \(1+x+x^2\)\label{abcas}
\een
A direct computation of the nonabelian Casimir energy is presently beyond reach. We will therefore approach this problem in an indirect by going through several steps as follows. First we will redo the abelian computation of the Casimir energy by using the correspondence between the Casimir energy on $\mb{R}\times S^5$ and the conformal anomaly on $S^6_q$. This computation can in turn be done in two different ways. We can perform a direct computation of the conformal anomaly on $S^6_q$ by completing the computation in \cite{Gustavsson:2019ick}. Alternatively, we can deduce what the Casimir energy shall be from just knowing the Renyi entropy on $S^6_q$ and the conformal anomaly on $S^6$ without $q$-deformation. The latter approach is the one we will use to compute the nonabelian Casimir energy.

We begin here by completing the abelian computation in \cite{Gustavsson:2019ick}. Since the abelian Renyi entropy was already obtained in \cite{Zhou:2015kaj} it will be enough for us to just spell out the Casimir energy terms, which are the terms that are linear in $q$ in the conformal anomaly, and put dots for the remaining terms. For one real conformal scalar, we have the conformal anomaly
\bea
a^S(q) &=& \frac{1}{15120 q^5} + \frac{1}{4320 q^3} + \frac{31}{30240} q
\eea
For one complex conformal scalar, we have twice of this result when $r_1 = r_2 = 0$. If we turn on $r_1$ that couples to this scalar (while this is neutral under $r_2$), then we get the conformal anomaly
\bea
a^{S_{1,+}}(q) &=& ... + \(\frac{31}{15120} - \frac{r_1^4}{72} + \frac{r_1^6}{180}\) q
\eea
A corresponding result holds for the other complex scalar that is charged under $r_2$,
\bea
a^{S_{2,+}}(q) &=& ... + \(\frac{31}{15120} - \frac{r_2^4}{72} + \frac{r_2^6}{180}\) q
\eea
By taking $r_1 = 1$ we reproduce the result in \cite{Gustavsson:2019ick},
\bea
a^{S_{1,+}}(q) &=& ... - \frac{19}{3024} q
\eea
For the fermion we get
\bea
a^F(q,r) &=& ....+\(- \frac{367}{24192} + \frac{3 r^2}{128} - \frac{5 r^4}{1152} + \frac{r^6}{5760}\)q
\eea
where $r$ stands for either one of $r_1\pm r_2$. When $r=0$ we reproduce the result in \cite{Gustavsson:2019ick}
\bea
a^F(q,0) &=& ... - \frac{367}{24192} q
\eea
The total contribution to the M5 brane conformal anomaly is $2 a_6^F$. When we turn on R-gauge fields this is generalized to  
\bea
a^F(q) &=& a_6^F(q,r_1+r_2) + a_6^F(q,r_1-r_2)
\eea
We get the result
\bea
a^F(q) &=& ... + \(\frac{31}{3780} - \frac{11 r_1}{180} + \frac{r_1^3}{9} - \frac{r_1^4}{36} - \frac{r_1^5}{30} - \frac{r_1^5}{30} + \frac{r_1^6}{90}\)q
\eea
The total M5 brane conformal anomaly is
\bea
a^{M5} &=& a^S + a^{S_{1,+}} + a^{S_{2,+}} + \frac{1}{2} a^B - a^F
\eea
The individual terms are rather messy 
\bea
a^S(q) &=& ... + \frac{31}{30240} q\cr
a^B(q) &=& ... + \frac{191}{1008} q\cr
a^{S_{1,+}}(q) &=& ... + \(\frac{31}{15120} - \frac{r_1^4}{72} + \frac{r_1^6}{180}\) q\cr
a_6^{S_{2,+}}(q) &=& ... + \(\frac{31}{15120} - \frac{r_2^4}{72} + \frac{r_2^6}{180}\) q\cr
a^F(q) &=& ... + \(\frac{31}{3780} - \frac{11 r_1}{180} + \frac{r_1^3}{9} - \frac{r_1^4}{36} - \frac{r_1^5}{30} - \frac{r_1^5}{30} + \frac{r_1^6}{90}\) q
\eea
but when we add these contributions and use $r_1 + r_2 = 1$, we obtain a simple result
\bea
a^{M5}(q) &=& ... + \frac{1}{12} \(1 + r_1 r_2 + r_1^2 r_2^2\) q
\eea

Let us now obtain this result in a different way that we will later generalize to the nonabelian case as well. We use the result for the abelian Renyi entropy \cite{Nian:2015xky}
\bea
S(q) &=& \frac{s_3}{q^3} + \frac{s_2}{q^2} + \frac{s_1}{q} + s_0
\eea
where
\bea
s_3 &=& \frac{r_1^2 r_2^2}{12}\cr
s_2 &=& \frac{r_1 r_2}{12} - \frac{r_1^2 r_2^2}{4}\cr
s_1 &=& \frac{1}{12} \frac{r_1^2 r_2^2}{4}\cr
s_0 &=& \frac{1}{2} - \frac{r_1 r_2}{12} - \frac{r_1^2 r_2^2}{12}
\eea
The conformal anomaly can be determined up to one undetermined coefficient $a_{-1}$ by using the the definition of the Renyi entropy 
\bea
S(q) &=& \frac{\ln Z(q) - q \ln Z(1)}{1-q}
\eea
together with the definition of the conformal anomaly
\bea
\ln Z(q) &=& a_6(q) \ln (\mu r)\cr
\Vol(H^5) &=& \ln\(\frac{r}{\eps}\)
\eea
where we shall put $\mu \eps = 1$. We then get the conformal anomaly as
\bea
a(q) &=& \frac{a_3}{q^3} + \frac{a_2}{q^2} + \frac{a_1}{q} + a_0 + a_{-1} q
\eea
where the coefficients are related to those of the Renyi entropy as
\ben
a_3 &=& s_3\cr
a_2 &=& s_2 - s_3\cr
a_1 &=& s_1 - s_2\cr
a_0 &=& s_0 - s_1\label{ar}
\een
In this way we find that 
\bea
a_3 &=& \frac{r_1^2 r_2^2}{12}\cr
a_2 &=& \frac{r_1 r_2}{12} - \frac{r_1^2 r_2^2}{3}\cr
a_1 &=& \frac{1}{12} - \frac{r_1 r_2}{12} + \frac{r_1^2 r_2^2}{2}\cr
a_0 &=& \frac{5}{12} - \frac{r_1 r_2}{12} - \frac{r_1^2 r_2^2}{3}
\eea
where $a_{-1}$ can not be determined this way. We will now determine $a_{-1}$ by noting that $a(1)$ is independent of $r_1$ and $r_2$ and is given by 
\bea
a(1) &=& \frac{7}{12}
\eea
which is the standard value for the conformal anomaly on $S^6$ that has been computed for $r_1 = r_2 = 0$. That this value for $a(1)$ must persist when $r_1$ and $r_2$ is nonzero is obvious since the R gauge field is proportional to $r_i (q-1)$ which vanishes when $q=1$. We then find that 
\bea
a_{-1} &=& \frac{1}{12} \(1 + r_1 r_2 + r_1^2 r_2^2\)
\eea
The Casimir energy is related as $-2 E_C = a_{-1}$ that gives
\bea
E_C &=& - \frac{1}{24} \(1 + r_1 r_2 + r_1^2 r_2^2\)
\eea

\section{The nonabelian Casimir energy}
A general formula for the nonabelian Renyi entropy on $S^6_q$ was proposed in \cite{Zhou:2015kaj}. The result can be expressed as
\bea
S(q) &=& \frac{s_3}{q^3} + \frac{s_2}{q^2} + \frac{s_1}{q} + s_0 
\eea
where
\bea
s_3 &=& \frac{x^2}{12} r + \frac{x^2}{12} d h^{\vee}\cr
s^2 &=& \(\frac{x}{12} - \frac{x^2}{4}\) r + \(\frac{x}{3} - \frac{x^2}{4}\) d h^{\vee}\cr
s^1 &=& \(\frac{1}{12} + \frac{x^2}{4}\) r + \(\frac{1}{3} + \frac{x^2}{4}\) d h^{\vee}\cr
s^0 &=& \(\frac{1}{2} - \frac{x}{12} - \frac{x^2}{12}\) r + \(1 - \frac{x}{3} - \frac{x^2}{12}\) d h^{\vee}
\eea
The corresponding conformal anomaly coefficients 
\bea
a_3 &=& \frac{x^2}{12} r + \frac{x^2}{12} d h^{\vee}\cr
a_2 &=& \(\frac{x}{12} - \frac{x^2}{3}\) r + \(\frac{x}{3} - \frac{x^2}{3}\) d h^{\vee}\cr
a_1 &=& \(\frac{1}{12} - \frac{x}{12} + \frac{x^2}{12}\) r + \(\frac{1}{3} - \frac{x}{3} + \frac{x^2}{3}\) d h^{\vee}\cr
a_0 &=& \(\frac{5}{12} - \frac{x}{12} - \frac{x^2}{12}\) r + \(\frac{2}{3} - \frac{x}{3} - \frac{x^2}{3}\) d h^{\vee}
\eea
are obtained by using (\ref{ar}). We determine the coefficient $a_{-1}$ by demanding that the conformal anomaly
\bea
a(q) &=& \frac{a_3}{q^3} + \frac{a_2}{q^2} + \frac{a_1}{q} + a_0 + a_{-1} q
\eea
at the point $q=1$ reproduces the known conformal anomaly on $S^6$ \cite{Gustavsson:2019efu, Cordova:2015vwa, Maxfield:2012aw} for any value of $x$,
\bea
a(1) &=& \frac{7}{12} r + \frac{4}{3} d h^{\vee}
\eea
In this way, we get
\bea
a_{-1} &=& \frac{1}{12} \(1 + x + x^2\) + \frac{1}{12} (2 + x)^2 d h^{\vee}
\eea
Using the correspondence with the Casimir energy on $S^5$ we get this Casimir energy as
\ben
E_C &=& -\frac{1}{24} (1+x+x^2) r - \frac{1}{24} (2+x)^2 dh^{\vee}\label{EC}
\een
It is easy to see that (\ref{EC}) agrees with (\ref{Bull1}).

For $SU(N)$ gauge group we may take the large $N$ limit and match with our gravity computation. By using (\ref{sun}) we get the conformal anomaly
\bea
a(q) &=& N^3 \(\frac{x^2}{12 q^3} + \frac{x-x^2}{3q^2} + \frac{2 - 2 x + 3 x^2}{6 q} + \frac{2 - x - x^2}{3} + \frac{4 + 4 x + x^2}{12} q\) + \O(N^2)
\eea
By then defining 
\bea
q &=& \frac{1}{1+b}
\eea
this anomaly takes the form 
\bea
a(q) &=& N^3 \frac{(4+2 b + b^2 x)^2}{12 (1+b)} + {\cal O}(N^2)
\eea
and the Casimir energy becomes 
\bea
E_C &=& - \frac{N^3(2+x)^2}{24} + {\cal O}(N^2)
\eea
These results exactly agree  in the large $N$ limit with what we got on the gravity side.

\subsection*{Acknowledgement}
DB and AG were
supported in part by
NRF Grant 2017R1A2B4003095.  DB was also
supported in part by Basic Science Research Program
through National Research Foundation funded by the Ministry of Education
(2018R1A6A1A06024977).


\appendix
\section{Supersymmetry conditions}
The supersymmetry parameter transforms as 
\bea
\eps &\rightarrow & e^{i \frac{q-1}{2} \sigma^3} \eps
\eea
under $\tau \rightarrow \tau + 2\pi$. So when $q\neq 1$, in order to preserve some of the supersymmetries, we may turn on R gauge field parameters $r_1$ and $r_2$ that corresponds to transforming $\eps$ by 
\bea
\eps &\rightarrow & g\eps
\eea
where
\bea
g &=& e^{\(r_1 \h\Gamma_{12}+ r_2 \h\Gamma_{34}\)t} 
\eea
Then the supersymmetries that satisfy the supersymmetry condition
\bea
P \eps &=& 0
\eea
where 
\bea
P &=& \frac{1}{2} \[1 + i \(r_1 \h\Gamma_{12} + r_2 \h\Gamma_{34}\) \sigma^3 \]
\eea
will be preserved. For this condition to be a projection, we need to satisfy
\bea
P^2 &=& P
\eea
We find that this condition amounts to 
\bea
2 r_1 r_2 \h\Gamma \eps &=& \(1-r_1^2-r_2^2\) \eps
\eea
where $\h\Gamma = \h\Gamma_{1234}$. If $r_1 r_2 \neq 0$, then the solutions are  
\bea
r_1 + r_2 &=& \pm 1
\eea
and
\bea
\h\Gamma \eps &=& \eps
\eea
or 
\bea
r_1 - r_2 &=& \pm 1
\eea
and 
\bea
\h\Gamma \eps &=& -\eps
\eea
If $r_1 r_2 = 0$, then solutions are either $(r_1,r_2) = (1,0)$ and $(r_1,r_2) = (0,1)$ and for none of these solutions do we get a Weyl projection on $\eps$, which means that these solutions correspond to points with enhanced supersymmetry.

\section{Volume of hyperbolic space}
Here we will compute the volume of the hyperbolic space $H^{d+1}$ with unit radius and the metric 
\bea
ds^2 &=& d\eta^2 + \sinh^2\eta \, d\Omega_d
\eea
The volume is
\bea
\Vol(H^d) &=& \Vol(S^d) \int_0^{\eta_0} d\rho \sinh^d \rho
\eea
where we cut off at $\eta = \eta_0$. For $d$ odd
\bea
\int_0^{\rho_0} d\rho \sinh^d \rho &=& \frac{1}{2^d} \sum_{k=0}^d \(\begin{array}{c}
d\\
k
\end{array}\) \frac{e^{(2k-d)\rho_0}-1}{2k-d}
\eea
For $d$ even, we have an additional contribution coming from $k=d/2$,
\bea
\frac{1}{2} \(\begin{array}{c}
d\\
d/2
\end{array}\) \eta_0
\eea
The regularization amounts to defining a small parameter
\bea
\eps &=& r e^{-\eta_0}
\eea
The volume is now defined by making a small-$\eps$ expansion and by removing the singular terms. For even $d$, this procedure results in the volume
\bea
\Vol(H^{d+1}) &=& \Vol(S^d) \frac{(-1)^{d/2}}{2^d} \(\begin{array}{c}
d\\
d/2
\end{array}\) \ln \(\frac{r}{\eps}\)
\eea
For $d+1=5$ we get
\bea
\Vol(S^4) &=& \frac{\pi^2 16}{6}
\eea
and 
\bea
\Vol(H^5) &=& \Vol(S^4) \frac{6}{16} \ln \(\frac{r}{\eps}\)\cr
&=& \pi^2 \ln \(\frac{r}{\eps}\)
\eea

\end{document}